\documentclass [12pt,preprint]{aastex}
\begin{document}

\title{The Velocity Field Around Groups of Galaxies}

\author{F.D.A. Hartwick}

\affil{Department of Physics and Astronomy, \linebreak University of Victoria, 
Victoria, BC, Canada, V8W 3P6}
\begin {abstract}
A statistical method is presented for determining the velocity field in the 
immediate vicinity of groups of galaxies using only positional and redshift 
information with the goal of studying the perturbation of the Hubble flow 
around groups more distant than the Local Group. The velocities are 
assumed to obey a Hubble-like expansion law, 
i.e. $V=H_{exp}R$ where the expansion rate $H_{exp}$ is to be determined. 
The method
is applied to a large, representative group catalog and evidence is found for 
a sub-Hubble expansion rate within two well defined radii beyond the virial 
radii of the groups. This result is consistent with that of Teerikorpi et al.
(2008) who found a similar expansion law around 3 nearby groups and extends 
it to a more representative volume of space.  
\end{abstract}


\section {Introduction}

Teerikorpi et al. (2008) (hereafter T08) have recently shown that galaxies 
in the immediate vicinity of 3 nearby groups of galaxies exhibit a 
Hubble-like expansion law at a slightly sub-Hubble expansion rate. The 
authors had at their disposal measured velocities and distances in 
order to arrive at this result. T08 interpret their result in terms of the 
effects of dark energy. The arguments for the dark energy interpretation can 
be found in a series of papers starting with Chernin et al. (2000).
Briefly, when a density perturbation which is to become a galaxy group 
collapses, it leaves behind an empty annulus between the original turn-around 
radius designated here $R_{ES}$, and its post-collapse zero-velocity radius 
which lies between the virial radius and $R_{v}$. $R_{v}$ is the radius where 
the gravity force of the group mass is balanced by the antigravity force of
the vacuum energy. Should a galaxy find itself in the region between $R_{v}$ 
and $R_{ES}$ either as a result of the chaos of virialization or by later 
ejection from the group, it may be accelerated by the effects of the vacuum 
energy. Since as these authors argue, the radial motion of the particles 
within this shell should be nearly the same as Newtonian motion its 
expansion velocity can be described by the vacuum Hubble constant $H_{v}=
\sqrt{\Omega_{\Lambda}}H_{o}$. Maccio et al.(2005) found from N-body 
simulations that the observed coldness of the local Hubble flow around the 
local group is consistent with the dark energy interpretation.

However recently Hoffman et al. (2008) and Martinez-Vaquero et
al. (2009) using sophisticated cosmological simulations claim that
dark energy should have no effect on the local dynamics. In
particular, these authors showed that a similar local Hubble flow was
also found in an open cosmological model making the need for a
$\Lambda$ model unnecessary locally. Further, Sandage (1986) has shown 
analytically that sub-Hubble 
flows can exist around groups in cosmologies without a $\Lambda$ term though 
of a different form from that found by T08. It is 
clearly of interest to confirm and extend the T08 result. In order to do so a 
new method for finding the velocity field is required since with
increasing distances the only observational data available are
positions and redshifts both for the groups and the field
galaxies. The goal here is to devise such a method and to apply it to
an existing group catalog in order to determine if the perturbations to the 
Hubble expansion law found by T08 apply to groups in a larger and more 
representative volume of space.

\section {Fitting the Velocity Field Around Galaxy Groups with a Hubble-like 
Expansion Law}

Determining the expansion velocity in terms of Hubble-like law ($V=H_{exp}R$) 
requires knowledge of the velocity with respect to the group center in the 
direction of expansion $V$, 
and the distance from the group center $R$. To demonstrate the essence of the 
method, assume for the moment that two projections of this expansion law 
are available (i.e. $\Delta V=Vcos(\theta)$ and $\Delta R=Rsin(\theta)$ in  
Fig 1 ). Let $y^{\prime}=V~cos(\theta)$ and $x^{\prime}=H_{exp}R~sin(\theta)$.
Then
\begin{equation}
y^{\prime}/x^{\prime}=\frac{V}{H_{exp}R}cos(\theta)/sin(\theta)
\end{equation}
Now take the average of both sides of this equation assuming that the number 
of groups 
is large and that the surrounding field galaxies are randomly distributed 
about these groups and denote the mean of the angular term as 
\begin{equation}
I_{n}=\frac{1}{4\pi}\int_{sphere}cos(\theta)/sin(\theta)d\omega
\end{equation}
after rearranging terms $H_{exp}$ is varied until the ratio
\begin{equation}
\overline{(y^{\prime}/x^{\prime})}/I_{n}=1=\overline{V/H_{exp}R}
\end{equation}
and the true expansion rate is obtained. In practice, instead of  
$V~cos(\theta)$ and $R~sin(\theta)$ which are not known, we do have access 
to the components $V~cos(\beta)$ and $R~sin(\beta)$ (Fig.1) so an 
intermediate step is required and is described below.

The observational data available consists of the angular positions and 
redshifts of the group centers and the field galaxies ($z_{g}$ and $z_{f}$). 
For a given cosmology, 
co-moving distances d(z) can be calculated from $d(z)=c/H_{o}\int_{0}^{z}dz^
{\prime}/((1-\Omega_{\Lambda})(1+z^{\prime})^{3}+\Omega_{\Lambda})^{1/2}$. 
Note that  
$\Omega_{k}=0$ is assumed allowing the use of Euclidean geometry. Further
the redshifts of all the data to be analysed are $\lesssim 0.1$ so that the 
changes due to different assumed cosmologies are small. From Fig.1 
with $d_{g}$ now available
\begin{equation}
r_{p}=R~sin(\beta)=d_{g}sin(\alpha)
\end{equation}
and
\begin{equation}
d_{per}=d_{g}cos(\alpha)
\end{equation}
Numerically inverting the above expression for co-moving distance allows the 
calculation of $z_{per}$, the redshift at $d_{per}$.

Now consider the line of sight projection of the expansion law  
$\Delta V_{D}$ where
\begin{equation}
\Delta V_{D}=V~cos(\beta)\simeq c(z_{f}-z_{per})
\end{equation}
and define the ratio $y/x$ as follows
\begin{equation}
y/x=\Delta V_{D}/(r_{p}~H_{exp})=(V/H_{exp}R)cos(\beta)/
sin(\beta)
\end{equation}
Rotate the line-of-sight about the observer by $+~\alpha$ and define a new 
ratio in terms 
of $\theta=\alpha+\beta$. Since $V$ and $R$ remain unchanged by this 
operation we obtain
\begin{equation}
y^{\prime}/x^{\prime}=\mid(-sin(
\alpha)+(y/x)cos(
\alpha))/(cos(\alpha)+(y/x)sin(\alpha))\mid=(V/H_{exp}R)\mid cos(\theta)/sin(
\theta)\mid
\end{equation}
Note that by taking the absolute value in (8) the integral over the sphere 
becomes by symmetry
\begin{equation}
I_{n}=\frac{2\pi\int_{0}^{\pi/2}\mid(cos(\theta)/sin(\theta))\mid sin(\theta)
d\theta}{2\pi\int_{0}^{\pi/2}sin(\theta)d\theta}=1
\end{equation}
Hence
\begin{equation}
\overline{V/H_{exp}R}=\overline{y^{\prime}/x^{\prime}}
\end{equation}
where the true value of $H_{exp}$ is that which makes the right hand side of
(10) equal to unity. For convenience we shall refer to the statistic 
represented by (10) as $Y(H_{exp})$.
Lastly from Fig 1 and equation (7)
\begin{equation}
R(H_{exp})=r_{p}/sin(\beta)=r_{p}/sin(cot^{-1}(\Delta V_{D}/(r_{p}H_
{exp})))
\end{equation}
where $H_{exp}$ is the value obtained above.

The above analysis applies to any one region undergoing Hubble-like expansion 
at a rate $H_{exp}$.

The region with the sub-Hubble expansion rate ($H_{v}$) found by T08 lies 
between $R_{v}$
and $R_{ES}$ in Fig.1. ($R_{v}$, $R_{ES}$ and $H_{v}$ are defined 
in the next section). Beyond $R_{ES}$ the expansion is 
assumed to return
to the global Hubble rate $H_{o}$ ($H(z)$). (At higher $z$ the 
global rate becomes $H(z)=H_{o}
((1-\Omega_{\Lambda})(1+z)^3+\Omega_{\Lambda})^{1/2}$). Similarly, by (11) in 
each of these regions we then have $R(H_{v})$ and $R(H(z_{g}))$.

With two expanding regions, the four key quantities in the analysis are the 
mean of the ratio $y^{\prime}/x^{\prime}$ (equ'n 10) evaluated with $H_{v}$ 
and 
$H(z_{g})$ replacing $H_{exp}$ designated here $Y(H_{v})$ and $Y(H(z_{g}))$ 
and the normalized distances 
$R(H_{v})/R_{v}$ and $R(H(z_{g}))/R_{v}$. These four quantities are calculated
for every field galaxy. In order to confirm the T08 result, $Y(H_{v})$ should 
be unity 
when $1 \leq R(H_{v})/R_{v} \leq R_{ES}/R_{v}$ for the particular value of 
$H_{exp}=H_{v}$ and $Y(H(z_{g}))$ 
should be unity when $R(H(z_{g}))/R_{v}$ becomes larger than 
$R_{ES}/R_{v}$. 
Both of the above conditions rely on the assumption that the galaxies are 
randomly distributed about the group centers for the sample as a whole. Note 
also that because we are dealing with ratios, the result is independent of the 
actual value assumed for $H_{o}$.

In order to validate the above procedure a simple numerical simulation
was performed. In this simulation the same group properties i.e. the
positions, redshifts and masses (from which $R_{v}$ could be
calculated) of groups used in the analysis of the observations were
used as centers around which `field galaxies' were randomly
distributed (both in angle $\theta$ and group-centric distance
$R$). Around each of the 12620 group centers are 300 randomly chosen
field galaxies. For a given cosmology, line-of-sight redshifts were
assigned (either as a result of global Hubble expansion or sub- Hubble
expansion about the group center) depending on whether $R/R_{v}$ was
greater than or less than 2. The simulation is clearly idealized to
the extent that each group and its retinue of randomly distributed
galaxies was treated as an isolated system. The data were then
analysed (as described above) using the same routines that were used
for the observations. Here the values of $R$ plotted on the abscissa come from 
the simulation. Shown in Fig. 2 are the results for an assumed
($\Omega_{m},\Omega_{\Lambda},\Omega_{k}$) (0.3,0.7 ,0.0)
cosmology. The red points represent $Y(H_{v})$ and $R/R_{v}$. The
black points were calculated assuming $H_{exp}=H(z_{g})$ and plotted
as $Y(H(z_{g}))$ versus $R/R_{v}$. Note the characteristic step in the
distribution of both sets of points at the transition point
($R/R_{v}=2$) and the continuity of red and black points at
$\overline{V/H_{exp}R}=1$.

\section{Analysis of the Observations}

\subsection{The Data}
The Tago et al. (2006) group catalogue of galaxies from the 2DF redshift 
survey (Colless et al. 2001, 2003) consists of 3 tables-one enumerates the 
groups and their properties, the 
second lists the group members while the third gives the co-moving distances 
of the galaxies which do not belong to any group. This catalogue was 
constructed using a friends-of friends algorithm. As such any individual 
group is subject to contamination problems (e.g. Niemi \& Valtonen 2009). In 
the application discussed here, the groups are used statistically so that 
any contamination effects will affect the determination of both sub-Hubble and
the global Hubble rate in the same way. For this 
investigation we limited our selection of groups to those lying within 300/h 
Mpc. This sample contains 12620 groups. The virial theorem
(i.e. $M=(\pi /2)r_{virial}3\sigma^{2}/G$) where $r_{virial}$ and $\sigma$ are
given in the first table of the group catalog was used to determine the mass 
of each of the groups. These masses were then used to calculate values of 
$R_{v}$ from equation (12) below.

For a given group each field galaxy subtends an angle $\alpha$ between it and 
the group center and each lies at the tabulated co-moving distance computed 
for a 
($\Omega_{m},\Omega_{\Lambda},\Omega_{k}$) (0.3,0.7,0.0) with $H_{o}=100h$ and
$h=1$ 
cosmology. These data are used to recover the individual redshifts ($z_{g}$ 
and $z_{f}$) which the authors had corrected for the motion relative to the 
CMB. The redshifts are then used to recompute the co-moving distances $d_{g}$ 
and $d_{f}$ (now assuming h=0.7) for the different trial cosmologies 
considered below.

\subsection{Some Definitions}

For the purpose of comparing the velocity field derived here with that of 
T08, we define the following analogous quantities.

Let $M$ represent the mass of an individual group and define the 
two radii $R_{v}$, and $R_{ES}$ as follows.
\begin{equation}
M/(4\pi R_{v}^{3}/3)=2\rho_{v}=2\Omega_{\Lambda}\rho_{c}
\end{equation}
and
\begin{equation}
M/(4\pi R_{ES}^{3}/3)=\rho_{m}=(1-\Omega_{\Lambda})\rho_{c}
\end{equation}
here $\rho_{c}$ is the critical density and is equal to $3H_{o}^{2}/8\pi G$. 
$R_{v}$ is the radius where the gravity force of mass M is balanced by the 
antigravity force of the dark energy \footnote{The factor of 2 
in equation (12) arises because in General Relativity the effective density 
includes a contribution from the pressure as well i.e. $\rho_{eff}=\rho + 3P/
c^{2}$ and $P=-\rho_{v}c^{2}$.}. $R_{ES}$ is the radius within which the 
the mass now in the group originated and beyond which the velocity field 
approaches the global Hubble flow (Teerikorpi \& Chernin, 2010 hereafter T10).
Note that 
$R_{ES}/R_{v}=(2\Omega_{\Lambda}/(1-\Omega_{\Lambda}))^{1/3}$ and is greater 
than one as long as $\Omega_{\Lambda}> 1/3$. 
The region within $R_{ES}$ is referred to by T08 as an  
Einstein-Straus vacuole. Also recall that $M/(4\pi R_{virial}^{3}/3)\sim 200
\rho_{c}$ so that $R_{v}/R_{virial}\sim 5.1$. Rearranging (12)
\begin{equation}
R_{v}=(GM/(H_{o}^{2}\Omega_{\Lambda}))^{1/3}
\end{equation}

Finally we parameterize the unknown sub-Hubble 
expansion rate as 
\begin{equation}
H_{v}=\sqrt{\Omega_{\Lambda}}H_{o}
\end{equation}
where following T08 we have neglected higher order terms (Chernin et al., 
2000). This expression (referred to
as the vacuum Hubble constant) also follows from the 
first Friedmann equation with $\Omega_{m}$ and $\Omega_{k}$ set to zero and is 
independent of redshift.

Before presenting the observations, we discuss some limitations of the method 
imposed by the observations themselves. Velocity components along the line of 
sight beyond the 
projection of the actual expansion velocity will bias our statistic to be 
larger than
one. Two contributors are redshift errors and peculiar motions (a  
`noisy' Hubble flow). Redshift errors are $\sim 85~kmsec^{-1}$ (Colless et al.
2001). The imprecise location of the group center will generally bias our 
statistic to be less than one.  If both the inner and global expansion laws 
are affected in the same way the severity of the above biases can be 
monitored by evaluating the statistic for galaxies undergoing global Hubble 
expansion. Another point to consider is that groups that are close 
together can have the same field galaxies in common. However, 
just as galaxies participating in the Hubble flow are doing so irrespective of
the co-moving center, the same argument applies to galaxies assumed to be 
expanding within the 
vacuum dominated region. For this reason, we choose galaxies 
within each region $1 < R/R_{v} < R_{ES}/R_{v}$ (vacuum dominated region) and 
$R/R_{v} > R_{ES}/R_{v}$ (global Hubble expansion region)
and treat them as separate samples. 

In order to determine the dividing line between sub-Hubble and global
expansion, solutions were obtained for the data broken up into small
groups with similar values of $R_{v}$. It was found that the distinct
step pattern seen in the simulation (Fig. 2) while visible for the
whole sample was most clearly defined when using only those groups
with $2.6 < R_{v}(h^{-1}_{70}Mpc) < 3.3$ for $\Omega_{\Lambda}=0.7$. 
The corresponding
range in group mass is $\sim 1-2 \times 10^{13}h^{-1} M_{\odot}$. The
results are shown in the top two panels of Fig. 3. Similar results
were found for other trial values of $\Omega_{\Lambda}$ after allowing for the 
$\Omega_{\Lambda}$ dependence of $R_{v}$ and are shown in
Fig. 5. The bottom panel of Fig. 3 shows the results of assuming an
open cosmology model (OCDM) with $\Omega_{m}=0.3,\Omega_{\Lambda}=0,
\Omega_{k} =0.7$ for the same groups and with the above values of $R_{v}$ 
used here in order to allow a comparison of results between the two 
cosmological models. Technically $R_{v}= \infty$ for the open model. Common to 
all 3 panels, the point-to-point scatter 
appears smaller below the break than it does above. The dispersion of the 
distribution of $Y(H(z_{g}))$ for 117896 group-galaxy pairs 
above the break (middle panel) is twice that of the dispersion of $Y(H_{v})$ 
for the 9894 group-galaxy pairs 
below the break (top panel). Comparable numbers apply to the other models 
shown in Fig. 5. If the galaxies below the break originated from
within the group they will have the group peculiar motion imprinted on
them and they will exhibit a `cooler' flow. Beyond the break the
galaxies are uncoupled from the groups and show the full effects of
peculiar motion. This applies here to both the OCDM and $\Lambda$CDM
model because the break appears in the same place. The bottom 2 panels
show that the global Hubble rate cannot account for expansion
velocities of the galaxies in the region below the break. The
$\Lambda$CDM model can provide the appropriate sub-Hubble rate (top panel).

Fig. 3 shows that the dividing line between the two velocity regimes
is reasonably sharp and occurs at $R/R_{v}=2.0 \pm 0.1$. We identify
this particular value of $R$ as $R_{ES}$. (Note from equ'ns (12) and
(13) the ratio $R_{ES}/R_{v}$ is predicted to be 1.7 when
$\Omega_{\Lambda}=0.7$ and 2 when $\Omega_{\Lambda}=0.8$ (T10).
A possible explanation for this small discrepancy 
comes from the observations of galaxy clusters that a substantial amount of 
mass (up to $100 \%$) exists outside the virial radius but within the post-
collapse zero-velocity radius which is below $R_{v}$ (e.g. Rines \& Diaferio 
2006). Allowing for this possibility would increase $R_{v}$ by $2^{1/3}$ and 
the predicted ratio would be recovered). Another possible contributor to the  
discrepancy in $R_{ES}/R_{v}$ is our neglect of the `lost-gravity' term due to
dark energy in the calculation of the virial mass (Chernin et al. 2009). It is 
instructive to divide the data 
at this point and to plot the statistic (equ'n 10) versus $R_{v}$
itself. The results are shown in Fig. 4 for the case
$\Omega_{\Lambda}=0.7$. The top panel shows results for $R/R_{v} < 2$
for all 12577 groups with $R_{v} < 5.7 h^{-1}_{70}Mpc$. The middle panel shows
results for the same sample for $R/R_{v} > 2$. As before the red
points represent the statistic calculated with
$H_{exp}=H_{v}=0.84H_{o}$ (i.e. $\Omega_{\Lambda} =0.7$) while the
black points were obtained with $H_{exp}=H(z_{g})$. The dotted lines
enclose the region used to construct Figs. 3 \& 5. The region at
$R_{v} < 2h^{-1}_{70}Mpc$ never reaches unity while the data become very noisy
at $R_{v} > 3.3h^{-1}_{70}Mpc$ and the break becomes less well defined. It is 
interesting that 
the N-body simulations of Wang et al. (2009) show (their Fig. 6) that for host
halo masses greater than $10^{13}h^{-1} M_{\odot}$ ejected sub-halos fall back
while for host halo masses lower than this ejected sub-halos are likely to 
reach the region beyond $R_{v}$. The data in 
this higher mass range are too sparse to determine whether this same 
phenomenon is responsible for the break becoming less well defined. Plots for 
higher and lower values of $\Omega_{\Lambda}$ are similar with the statistic
among the red points between the dashed lines in the $R/R_{v} < 2$
plots asymptoting at lower and higher values. The results for $R/R_{v} > 2$
(black points) for a range in vacuum energy are very similar
to those shown in the middle panel. It is not clear what is
responsible for the fall-off below $R_{v} \sim 2h^{-1}_{70}Mpc$ in these plots
especially given that T08 found their result at $R_{v} \sim 1.3h^{-1}_{70}
Mpc$.  A
possible explanation comes from the observation that the galaxies that
are supposed to be expanding at the Hubble rate (those plotted in the
middle panel) also show large deviations from unity in the same low
range of $R_{v}$ suggesting that the field galaxies surrounding these
low mass groups may not be uniformly distributed. 

A sub-sample of more isolated groups was analysed in order to measure
the effect of group clustering on the result. This sub-sample included
only those 3894 groups of the original 12620 without a neighbor within
$4h^{-1}_ {70}Mpc$. The results are shown in the bottom panel of
Fig. 4 and can be compared to the top panel. The previous results
shown in Figs 3 \& 5 remain unchanged but with larger uncertainty. In
addition the fall-off below $R_{v} \sim 2h^{-1}_{70}Mpc$ noted above is still
present. This region should show the most change if the effect of 
clustering is a dominating influence. Increasing the isolation criterion 
above leaves too few groups and very poor statistics.

As a further check on the results another sub-sample consisting of only those
1835 groups with more than 5 members was analysed. These results have 
larger uncertainties but there are no systematic differences from those shown 
in Figs 3 \& 4.

The main results of this work are summarized in Fig. 5 which shows how
the solutions vary as a function $\Omega_{\Lambda}$ or degree of 
sub-Hubble expansion. The statistic in
the individual panels in Fig. 5 was determined in exactly the same
manner as that in the simulation in Fig. 2. Red points represent
$Y(H_{v})$ and black points $Y(H(z_{g})$. To the left of the break at
$R/R_{v} \sim 2$ these ordinates are plotted against $R(H_{v})/R_{v}$
while to the right against $R(H(z_{g})/R_{v}$. The same groups were used to 
construct each panel. These included those with $2.6 < R_{v} (h^{-1}_{70}Mpc)
<3.3$ in the 
middle panel. The large scatter, particularly among the black
points, is generally to values greater than unity. The most deviant
among these points also tend to show the largest uncertainties. As
discussed earlier, the full effects of the peculiar motions are
assumed to be causing this bias. The red points represent a region
with a cooler flow so that this bias should be negligible here. Giving
most weight to the red points, we adopt the value of $\Omega_{\Lambda}
\sim 0.7$ and based on the spread of the mean of the red points from
each panel assign an uncertainty of $\pm 0.1$. Using the more direct
approach T08 find $\Omega_ {\Lambda} \sim 0.77$ from known distances
and velocities. Finally we note from Fig. 1 of T10 that the actual value of 
$H_{v}$ should increase slightly over the range $1 < R(H_{v})/R_{v} < 1.7$.

\section{Summary and Conclusions}

From a combination of geometry and statistics we have shown that the velocity 
field surrounding groups of galaxies obeys 
two Hubble like expansion laws i.e. $V=H_{exp}R$. At large distances the 
global Hubble Law is recovered. Within two well defined radii $R_{v}$ and 
$R_{ES}$ we find a lower expansion rate ($H_{exp} \sim 0.84H_{o}$). Thus the 
main result of this work is the demonstration that the phenomenon reported by 
T08 extends well beyond
the vicinity of the local group. If one adopts the the interpretation of T08 
that the result is due to the local effects of dark 
energy, we find $\Omega_{\Lambda}\sim (0.84)^{2}
\sim 0.7 \pm 0.1$ from a relatively small range of group mass. It is 
interesting that Sandage (1986), using the data then available,
found that the calculated deceleration due to the Local Group is equivalent to 
$H_{exp} \sim 0.83H_{o}$ at a distance $ \sim 1.6h^{-1}_{70}Mpc$ for an 
$\Omega_{0}=0$ $(\Lambda=0)$ cosmology.

These results do not contradict the conclusions from the cosmological
simulations of Hoffman et al. (2008) \& Martinez-Vaquero et al. (2009)
that the `coolness' of the local flow, by itself is not a strong
discriminant between a model without $\Lambda$ (OCDM) and
$\Lambda$CDM. Here the flow between $R_{v}$ and $R_{ES}$ is cooler
because the galaxies within this region originated from inside the
group and hence share the group's peculiar motion. Galaxies beyond
$R_{ES}$ feel the full effect of the peculiar motions.  The
observations show a systematic lowering of the expansion rate 
within the inner (cool flow) region in both of the OCDM and
$\Lambda$CDM models. Based on the T08 
interpretation the sub-Hubble rate in the $\Lambda$CDM model is predictable 
and when applied can account for the observations. However, interpreting the 
results as a local cosmological effect $does$
contradict the conclusions from the above simulations. Until this impasse
is definitively resolved with further work one may consider $H_{v}$ to
be a convenient parameterization of a sub-Hubble expansion rate.

In conclusion, significant perturbations to the global flow around galaxy 
groups beyond the Local Group are present. The sub-Hubble expansion rates are 
similar to those found recently by T08 around 3 nearby groups. The possibility 
that {\it{local}} dark energy effects are responsible makes the problem worthy
of further investigation both observationally and theoretically.

\section{Acknowledgements}

The author acknowledges helpful discussions in the early stages of 
this work with Tony Burke and Chris Pritchet and financial support from an 
NSERC Canada Discovery grant. In addition the author thanks the referee for 
providing a very constructive report. He also gratefully acknowledges guest 
worker privileges at the Herzberg Institute of Astrophysics.

\clearpage

\begin{figure}
\plotone{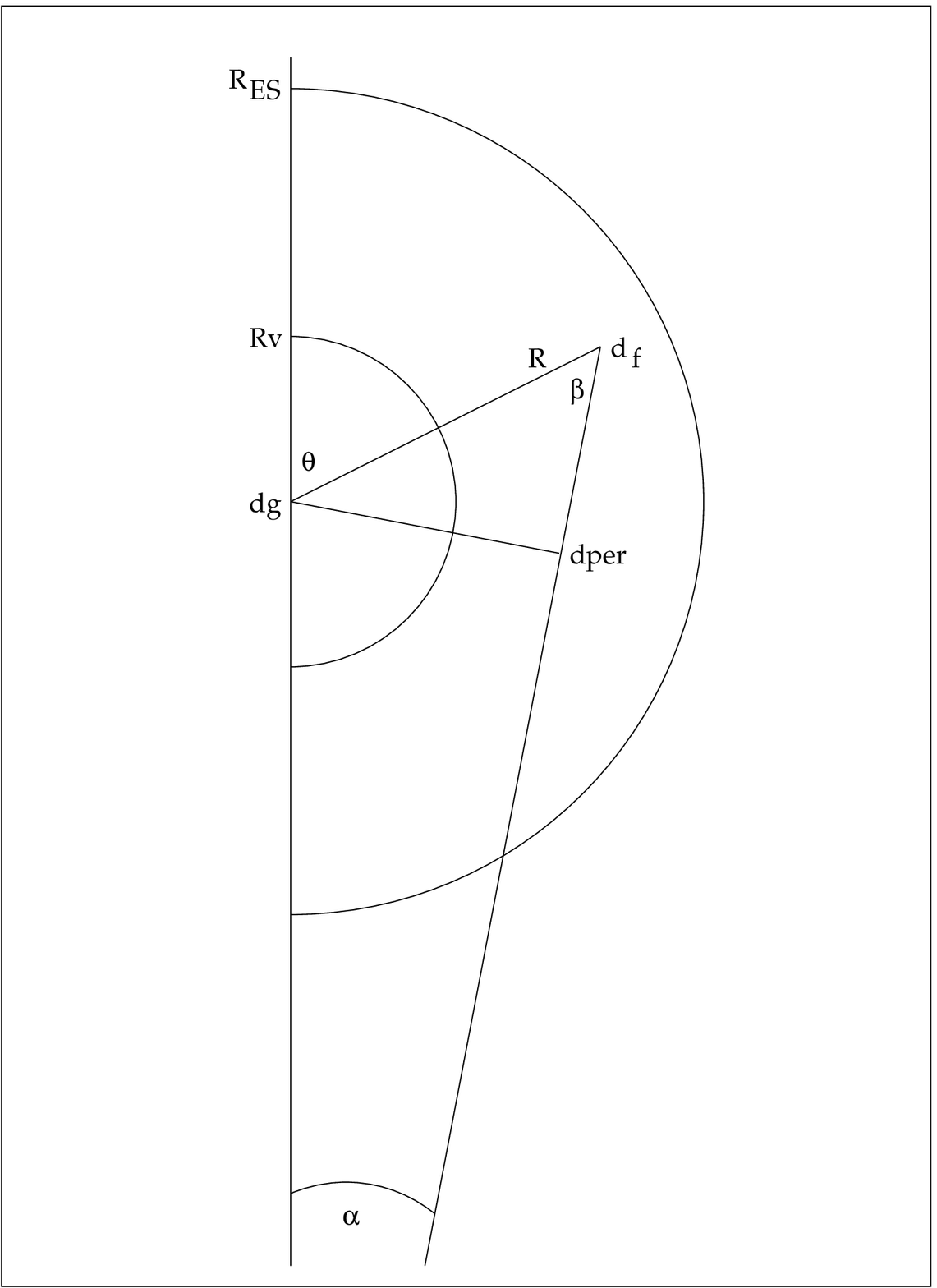}
\figcaption{The geometry of the problem is shown. The group is at distance 
d$_{g}$ while the field galaxy at $d_{f}$ subtends an 
angle $\alpha$ at the observer. See text for details.}
\end{figure}

\begin{figure}
\plotone{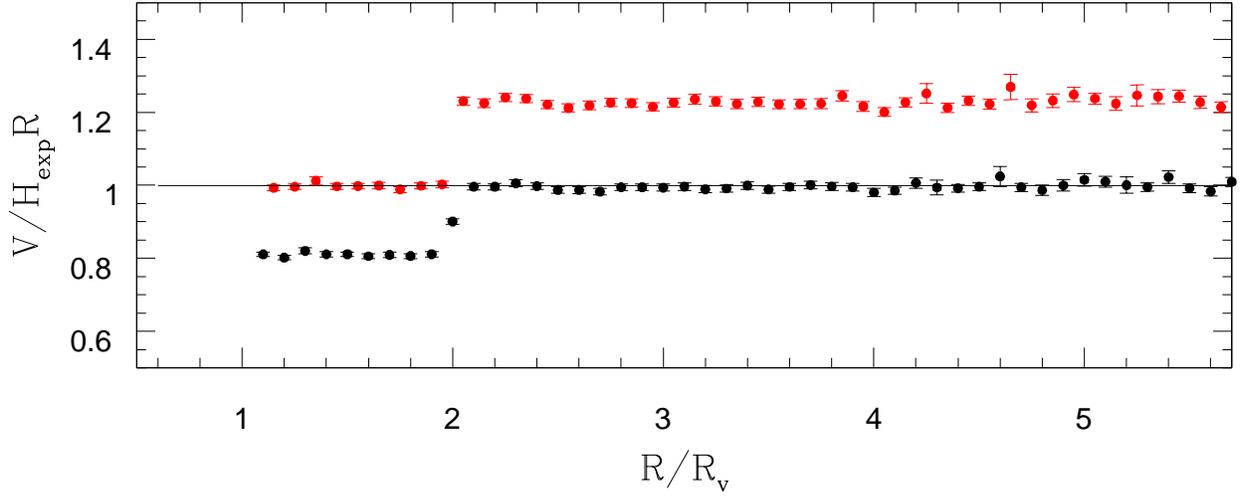}
\figcaption{The results of a simulation showing the average $V/H_{exp}R$ 
versus normalized 
distance from the group center $R/R_{v}$. In this simulation group properties 
were the same as used in the analysis of the observations of the field 
galaxies. Here 
the `field galaxies' were assigned random angles of $\theta$, assigned 
random 
group-centric distances $R$ and given appropriate line of sight redshifts due 
either to global 
Hubble or sub-Hubble expansion about these centers if $R/R_{v}$ was greater 
than or less than 2. Three hundred field galaxies were randomly assigned to 
each of the 12620 groups. The result shown is for an ($\Omega_{m}=0.3,\Omega_
{\Lambda}=0.7,\Omega_{k}=0$)
cosmology. The red points then represent $Y(H_{exp})$ with $H_{exp}=H_{v}=
0.84H_{o}$ while the black points show results with $H_{exp}=
H(z_{g})$.  Error bars here and in all of the 
following diagrams represent the standard error of the mean.}
\end{figure}

\begin{figure}
\plotone{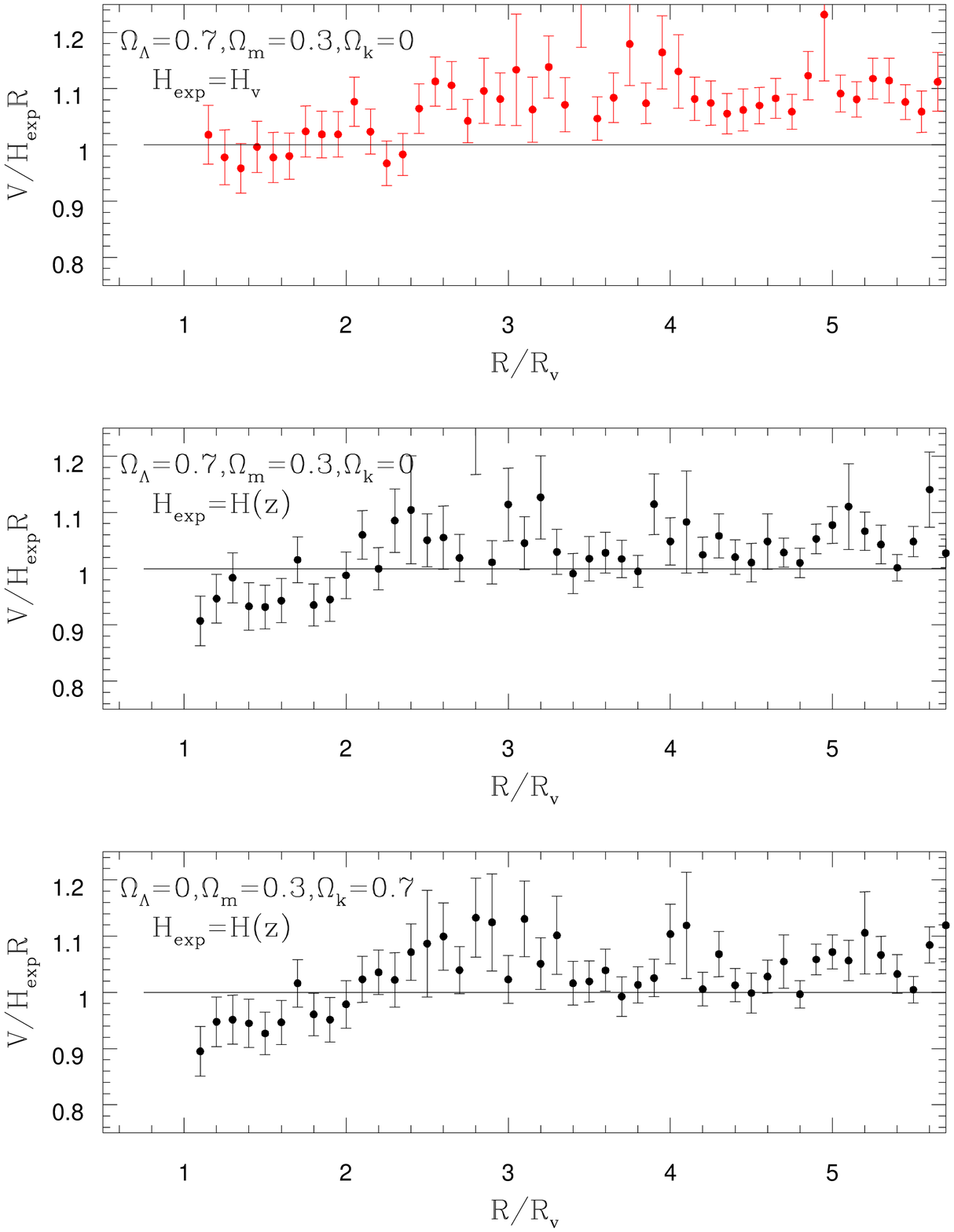}
\figcaption{Average $V/H_{exp}R$ plotted against $R/R_{v}$ for the cosmology 
and the assumed expansion velocity indicated. Only groups with 
$2.6 < R_{v}(h^{-1}_{70}Mpc)
 < 3.3$ were included as it is within this interval that the break
at $R/R_{v} \sim 2$ was most distinct. This point represents the transition 
from sub-Hubble to global Hubble expansion ($R_{ES}$). The bottom panel shows 
analogous results for an open cosmology. See the text for a discussion of the 
scatter in these plots. Note in the top panel how the $\Lambda$ model can 
account for the expansion velocities below the break.} 
\end{figure}

\begin{figure}
\plotone{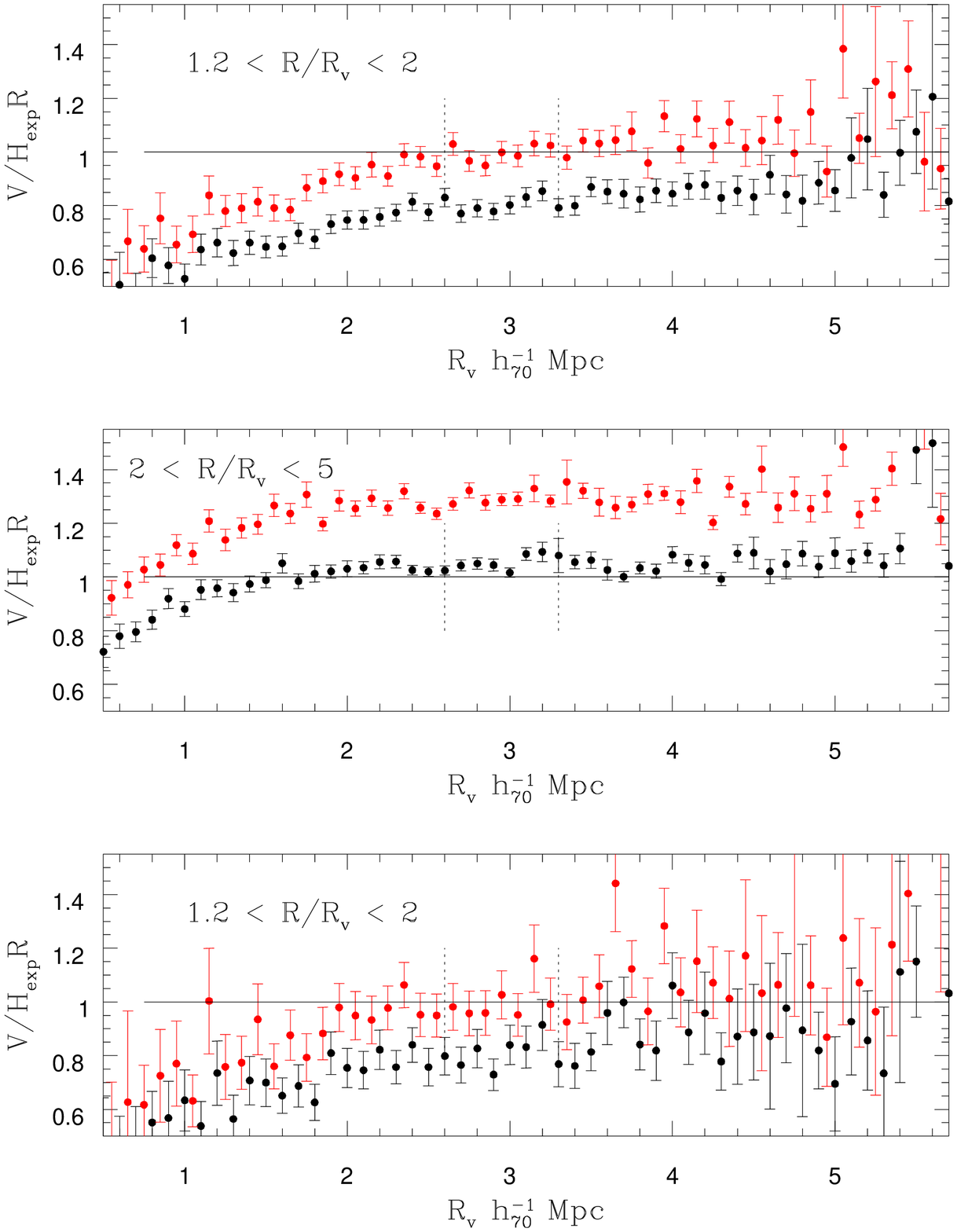}
\figcaption{Top and middle panels: Average $V/H_{exp}R$ plotted against 
$R_{v}$ for all 12577 
groups with $R_{v} < 5.7$ and for the two regions of $R/R_{v}$ indicated. 
As in the other figures 
red points represent $Y(H_{v})$ and black points $Y(H(z_{g}))$. The bottom 
panel is similar to the top but for a sub-sample of groups with no neighbor 
within $4h^{-1}_{70}Mpc$. If group clustering was influencing the results the  
largest difference would occur between $R_{v}$ of 1 and 2. The dotted lines 
indicate the region within which groups were used to construct Figs. 3 \& 5.} 
\end{figure}

\begin{figure}
\plotone{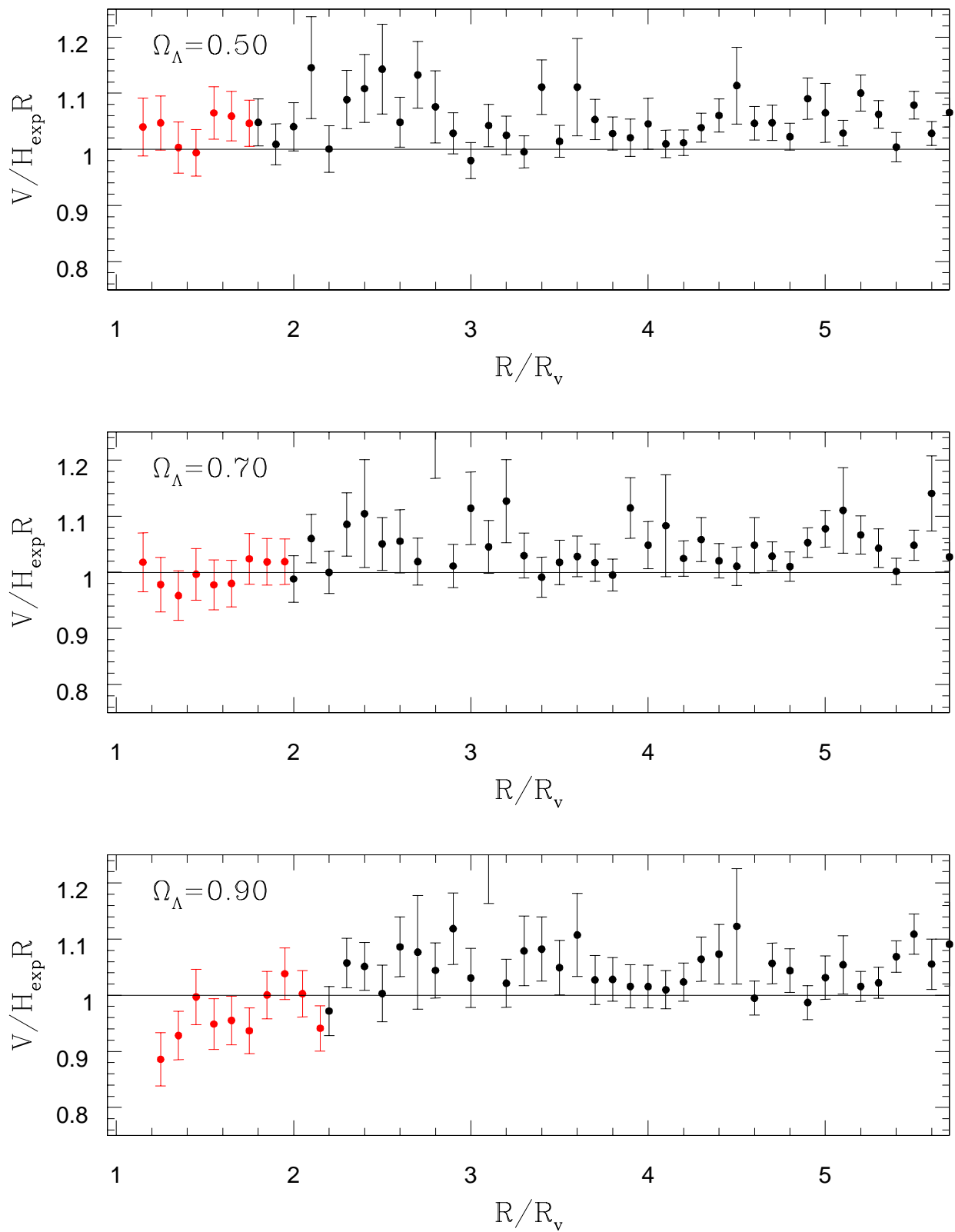}
\figcaption{The same as Fig.2 except now showing the results from an analysis 
of the group and field galaxy observations for three trial values of 
$\Omega_{\Lambda}$. As in the other figures 
red points represent $Y(H_{v})$ and black points $Y(H(z_{g}))$. Note that  
$R_{v}$ itself is dependent on the assumed value of $\Omega_{\Lambda}$.}
\end{figure}

\end{document}